\documentclass[%
 aip,
 amsmath,amssymb,
 reprint,%
]{revtex4-1}

\usepackage{color}
\usepackage{amsmath}
    \DeclareMathOperator{\sech}{sech}
\usepackage{pifont}% http://ctan.org/pkg/pifont
\usepackage{amssymb}  % More symbols
\usepackage{bbold}
\usepackage{float}
\usepackage{placeins}
\usepackage{tikz}
\usepackage{makecell}
\usepackage{subfigure}
\usepackage{scalerel}
\usepackage{pifont}   % Ding symbolshttps://www.overleaf.com/project/6084801cec8058e65885c58b
\usepackage{dcolumn}  % Align table columns on decimal point
\usepackage{multirow} % Table functions
\usepackage{placeins}% For making floats not move around everywhere
\usepackage{soul}
\usepackage{xcolor}
\usepackage{graphicx}% Include figure files
\usepackage{dcolumn}% Align table columns on decimal point
\usepackage{bm}% bold math
\usepackage{listings}

\newtoks \wid

\makeatletter
\def\@email#1#2{%
 \endgroup
 \patchcmd{\titleblock@produce}
  {\frontmatter@RRAPformat}
  {\frontmatter@RRAPformat{\produce@RRAP{*#1\href{mailto:#2}{#2}}}\frontmatter@RRAPformat}
  {}{}
}%
\makeatother
\begin{document}
\wid{0.45\textwidth}

\preprint{AIP/123-QED}

\title{Magneto-Optical Analysis of Magnetic Anisotropy in Ultrathin Tm$_{3}$Fe$_{5}$O$_{12}$/Pt Bilayers}
% Force line breaks with \\
\author{T. Nathan Nunley}
\affiliation{Department of Physics, Center of Complex Quantum Systems, The University of Texas at Austin, Austin, Texas 78712 USA  
\looseness=-1}
\affiliation{Center for Dynamics and Control of Materials and Texas Materials Institute, The University of Texas at Austin, Austin, Texas 78712, USA  
\looseness=-1}

\author{Daniel Russell}
\affiliation{Department of Physics, The Ohio State University, Columbus, Ohio 43210, USA  
\looseness=-1}

\author{Liang-Juan Chang}
\affiliation{Institute of Space Systems Engineering, National Yang-Ming Chiao-Tung University, Hsinchu 30010, Taiwan
\looseness=-1}

\author{David Lujan}%
\affiliation{Department of Physics, Center of Complex Quantum Systems, The University of Texas at Austin, Austin, Texas 78712 USA  
\looseness=-1}
\affiliation{Center for Dynamics and Control of Materials and Texas Materials Institute, The University of Texas at Austin, Austin, Texas 78712, USA  
\looseness=-1}

\author{Jeongheon Choe}
\affiliation{Department of Physics, Center of Complex Quantum Systems, The University of Texas at Austin, Austin, Texas 78712 USA  
\looseness=-1}
\affiliation{Center for Dynamics and Control of Materials and Texas Materials Institute, The University of Texas at Austin, Austin, Texas 78712, USA  
\looseness=-1}

\author{Side Guo}
\affiliation{Department of Physics, The Ohio State University, Columbus, Ohio 43210, USA  
\looseness=-1}

\author{Shang-Fan Lee}
\affiliation{Institute of Physics, Academia Sinica, Taipei, 11529, Taiwan
}

\author{Fengyuan Yang}
\email{yang.1006@osu.edu}
%\altaffiliation{Authors to whom correspondence should be addressed. elaineli@physics.utexas.edu, yang.1006@osu.edu}
\affiliation{Department of Physics, The Ohio State University, Columbus, Ohio 43210, USA  
\looseness=-1}

\author{Xiaoqin Li}
\email{elaineli@physics.utexas.edu}
%\altaffiliation{Authors to whom correspondence should be addressed. elaineli@physics.utexas.edu, yang.1006@osu.edu}
\affiliation{Department of Physics, Center of Complex Quantum Systems, The University of Texas at Austin, Austin, Texas 78712 USA  
\looseness=-1}
\affiliation{Center for Dynamics and Control of Materials and Texas Materials Institute, The University of Texas at Austin, Austin, Texas 78712, USA  
\looseness=-1}

\date{\today}% It is always \today, today,

\begin{abstract}
Magnetic bilayers consisting of an epitaxially grown ferrimagnetic insulator and a heavy metal layer are attractive for spintronic application because of the opportunity for electric control and read-out of spin textures via spin orbit torque. Here, we investigate ultrathin thulium iron garnet (TmIG)/Pt bilayers when the TmIG layer thickness is 3 nm and below using a sensitive Sagnac magneto-optical Kerr effect technique. We compare the hysteresis loops from out-of-plane and in-plane applied magnetic fields. The preferred magnetization orientation evolves with the TmIG thickness and the presence of the Pt overlayer.  We quantify the evolution of the magnetic anisotropy in these ultrathin films and find a significant change even when the TmIG thickness is varied by less than 1 nm. In these ultrathin films,  the presence of a Pt overlayer changes the effective anisotropy field by more than a factor of 2, suggesting that the interfacial anisotropy at the Pt/TmIG interface plays a critical role in this regime.
\end{abstract}

\maketitle

\section{Introduction}

Many ferrimagnetic insulators exhibit Curie temperatures significantly above room temperature. In their multilayer form, these materials provide promising platforms for both fundamental studies of magnetism ~\cite{back20202020,neubauer2009topological,maccariello2018electrical,raju2021colossal,roessler2006spontaneous,shao2018role,avci2017current,avci2017fast, avci2019interface,velez2019high} and future spintronic applications~\cite{manchon2019current,avci2021current,kim2022ferrimagnetic,stupakiewicz2021ultrafast,jiang2017skyrmions,everschor2018perspective,dohi2021thin}. For example, skyrmions are more readily stabilized in two-dimensional system than the 3D skyrmion crystal phase \cite{banerjee2014enhanced}. Epitaxially grown magnetic garnet thin films, e.g. TmIG (Tm$_{3}$Fe$_{5}$O$_{12}$) are particularly interesting as they may host nanoscale skyrmions stable at room temperature \cite{shao2019topological,ahmed2019spin,lee2020probing,nunley2022quantifying}. These films exhibit low damping, high crystalline quality with low pinning density~\cite{wang2013large,velez2019high,shao2018role,chanda2024temperature}, tunable interfacial Dzyaloshinskii-Moriya interaction (DMI) \cite{caretta2020interfacial,velez2019high,lee2020probing}, and strain-tunable anisotropy ~\cite{ahmed2019spin,lee2020investigation,lee2020crystal,shao2019topological,wang2014antiferromagnonic,cheng2020nonlocal, tang2016anomalous}. Many of these features are highly desirable for the formation and current-driven motion of chiral domain walls~\cite{avci2017fast,avci2019interface,velez2019high} and skyrmions.

    \begin{figure}[t]
        \centering
        \includegraphics[width=\the\wid]{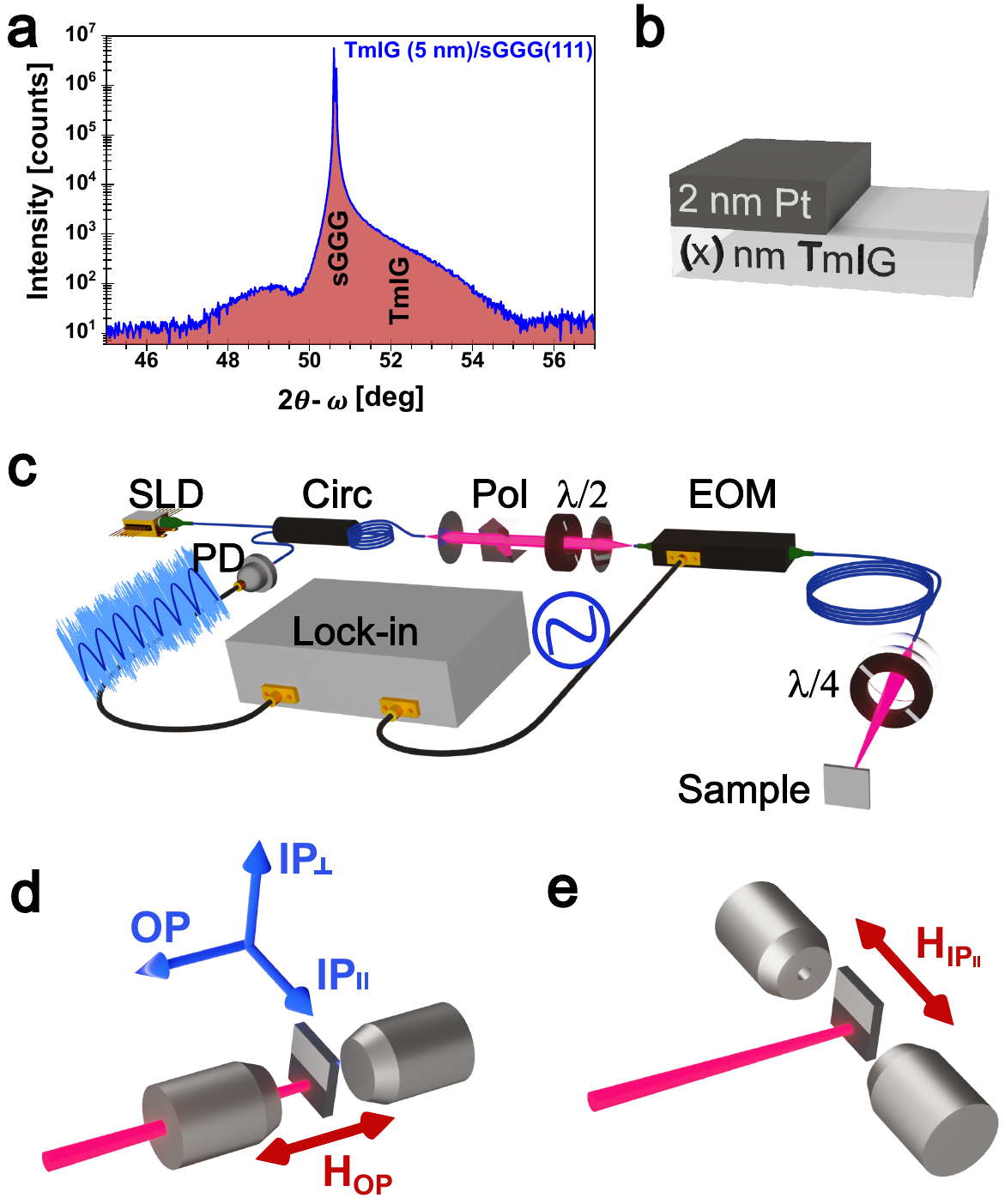}
        \caption{Sample quality and illustrations of experimental apparatus and sample structure. (a) is XRD of an epitaxially grown 5 nm TmIG film grown by the same method indicating the high crystalline quality of the TmIG film. (b) Rendering of the Pt/TmIG films geometry. (c) is the SaMOKE apparatus. The SLD is the superluminescent diode, Pol is the polarizer, $\lambda/2$ is the $\lambda/2$-waveplate, EOM is the electro-optic modulator, and $\lambda/4$ is the $\lambda/4$-waveplate. The lock-in measures the first and second harmonic of the reference frequency that is sent to the EOM. (d) and (e) Show the sample with the probe beam and electromagnet poles. The directional axes in (d) show the OP, IP$_{\parallel}$, and Transverse (IP$_{\perp}$) directions with respect to the sample surface. The IP$_{\perp}$ direction is perpendicular to the TmIG to Pt/TmIG transition. (d) shows the magnet and sample configuration for OP field, (e) for IP. }
        \label{fig:app}
    \end{figure}

Multiple parameters are important for generating robust skyrmions including saturation magnetization, exchange constant, DMI, and anisotropy. These parameters control the skyrmion type, size, and stability. The saturation magnetization and exchange constants are difficult to engineer because these properties are largely intrinsic. We note that some thickness dependence has been found, especially at the ultrathin limit \cite{shao2019topological,ahmed2019spin}. However, for ferrimagnetic films such as TmIG, anisotropy and DMI are highly tunable \cite{rosenberg2021magnetic,vu2020tunable,lee2020engineering,avci2019interface,caretta2020interfacial,ding2020identifying,ye2022spin,lee2020probing,xia2020interfacial,lee2020interfacial,ding2020identifying,xu2022strain}. A commonly used strategy is to choose different heavy metal layers. There have been extensive studies on engineering magnetic anisotropy of TmIG thin films through substrate-induced epitaxial strain \cite{rosenberg2021magnetic,vu2020tunable,lee2020engineering,lee2020crystal}, which leads to perpendicular magnetic anisotropy (PMA), However, debates continue about the role of a heavy metal overlayer and whether the interfacial DMI originate from the epitaxial interface or the insulator/metal interface \cite{caretta2020interfacial,lee2020interfacial,lee2020investigation,fakhrul2023influence}. The latter produces the spin Hall-topological Hall resistivity when topological spin textures in the bilayer form. Previously, the dependence of interfacial anisotropy at the insulator/metal interface on the crystal orientation has been investigated when the TmIG and YIG films are 5 nm thick \cite{lee2020crystal}. Below this thickness, traditional techniques, such as vibrating
sample magnetometry or superconducting quantum interference device magnetometry, do not offer sufficient sensitivity, partly because the very strong paramagnetic background from garnet substrates containing rare earth elements. Studying films less than 3 nm thick, the ultrathin limit, is critical for understanding topological magnetic phases near room temperature \cite{ahmed2019spin,nunley2022quantifying}. 

Here, we investigate how magnetic anisotropy in epitaxial TmIG bare films and Pt/TmIG bilayers evolves as a function of the film thickness in the ultrathin film limit. We apply a highly sensitive magneto-optical Kerr effect (MOKE)
technique to measure the hysteresis loops of TmIG films 2.3 to 3 nm thick. We compare the measurements when an out-of-plane (OP) vs. in-plane (IP) magnetic field is scanned. By extracting and comparing the saturation fields in these two geometries, we quantify the effective anisotropy field. In this ultrathin film limit, the interfacial anisotropy energy density becomes large compared to the volumetric anisotropy terms. We confirm that the Pt/TmIG interface produces a compass anisotropy that dominates the magnetic anisotropy when the TmIG film thickness is varied by less than 1 nm. Our study provides previously inaccessible insight into anisotropy engineering in these ultrathin ferrimagnetic insulating films and bilayers.

\section{Methods}

Epitaxial-Pt/TmIG bilayers were deposited by off-axis sputtering on Gd$_{2.6}$Ca$_{0.4}$Ga$_{4.1}$Mg$_{0.25}$Zr$_{0.65}$O$_{12}$ (sGGG) (111) substrates from MTI Corporation. The different TmIG and sGGG lattice constants, 12.324~\cite{tang2016anomalous,ahmed2019spin} and 12.480 \AA, ~respectively, result in a tensile strain which produces perpendicular magnetic anisotropy \cite{rosenberg2021magnetic,vu2020tunable}. An x-ray diffraction (XRD) scan of a representative 5 nm TmIG film grown on sGGG is shown in Fig.~\ref{fig:app}a with broad Laue oscillations around the TmIG(444) peak, demonstrating the high quality of the epitaxial film. It has been shown previously by some of us that this growth method produces high quality, single crystal films \cite{ahmed2019spin,lee2020probing,lee2020interfacial,lee2020crystal} without a magnetic dead layer at the TmIG/sGGG interface~\cite{lee2020crystal}. In order to explore the anisotropy transition, we prepared TmIG films of three different thicknesses, 3.0, 2.7, and 2.3 nm. After the TmIG growth, 2 nm Pt was deposited on each sample at room temperature. The TmIG films were partially blocked during the Pt deposition, leading to Pt coverage on only part of the sample surface. This method allows us to isolate the effect the Pt layer has on each TmIG film while keeping constant all other variables for each sample. The partial bilayer design is depicted in Fig. \ref{fig:app}b.

We apply a special magneto-optical Kerr effect (MOKE) technique known as the zero-area Sagnac MOKE (SaMoke)  to investigate these ultrathin films. This apparatus is designed to measure only polarization rotations arising from broken time-reversal symmetry \cite{deak2012reciprocity,rowe2017polarizers,armitage2014constraints,kapitulnik2015notes,vernon1980extension,potton2004reciprocity,xia2008optical,dodge1997sagnac}, allowing us to extract nrad-order polarization rotation in the Kerr angles arising from the reflection of the probe light from the sample surface. The SaMOKE setup is depicted in Fig. \ref{fig:app}c. Measurements were taken with 1550 nm probe light produced by a superluminescent diode at normal incidence in ambient conditions. The system is based on single mode polarization-maintaining fiber, with the reflected light coupling back into the fiber and traveling back through the system before being sent to the detector. The single mode fiber ensures strict adherence to electromagnetic reciprocity, allowing us to cancel all optical effects arising from unbroken reciprocity. The birefringence of the fiber is employed to create a polarization based same-path interferometer with zero residual area. The in-line fiber electro-optical phase modulator modulates the light before and after reflection with the modulation frequency set to achieve maximum double-modulation depth, determined by twice the optical path length from the modulator to the sample. The photodiode signal is demodulated at the first and second harmonics of the modulation frequency. Once the interferometer is calibrated, the ratio of the two harmonics can be used to calculate the Kerr angle, $\phi_{K} = \frac{1}{2}\arctan{([J_{2}(2\phi_{0})V_{\omega}]/[J_{1}(2\phi_{0})V_{2\omega}])}$, where $V_{\omega}$ and $V_{2\omega}$ are the demodulated harmonics, $J_1$ and $J_2$ are Bessel functions of the first kind, and $\phi_{0}$ is the maximum optical phase difference between light traveling in the fast and slow axes of the modulator. $\phi_{0}$ is proportional to the amplitude of the voltage applied to the modulator and is chosen to maximize $J_{1}$.

Polar MOKE at normal incidence is only sensitive to the OP components of the sample magnetization. As such, it is usually only used to measure systems for which the applied field is OP of the sample. Systems in which the domain orientation during rotation is randomized produce transient net magnetizations which are not easily analyzed. However, if the domains rotate coherently or there are few domains, we are able to easily interpret the OP component and correlate that to saturation magnetization in the IP$_{\parallel}$ direction. 

The major hysteresis loop of each sample were measured. The applied magnetic field was swept along the out-of-plane (OP) direction (Fig. \ref{fig:app}d) for all samples and the in-plane (IP$_{\parallel}$) (Fig. \ref{fig:app}e) for the 3.0 and 2.7 nm TmIG samples. Each sample in the OP and IP$_{\parallel}$ configurations was measured on both regions with a Pt overlayer and bare TmIG with the incident light focused in the middle of each respective region. The data taken with an OP field sweep and an IP field sweep are shown in Fig.~\ref{fig:OP} and Fig.~\ref{fig:ip}, respectively. A paramagnetic linear background is subtracted after fitting the data with a sum of hyperbolic and a linear functions. The paramagnetic background arises primarily from the sGGG substrate and Pt film contributions. All datasets with the paramagnetic background are shown in the Supplementary Information. Each data set is the average of 200 or more field sweeps in order to achieve an average fit residual of less than 10 nrad for the OP sweeps.

\section{Results}

\subsection{Out of Plane Field Sweeps}

\begin{figure}[t!]
        \centering
        \includegraphics[width=\the\wid]{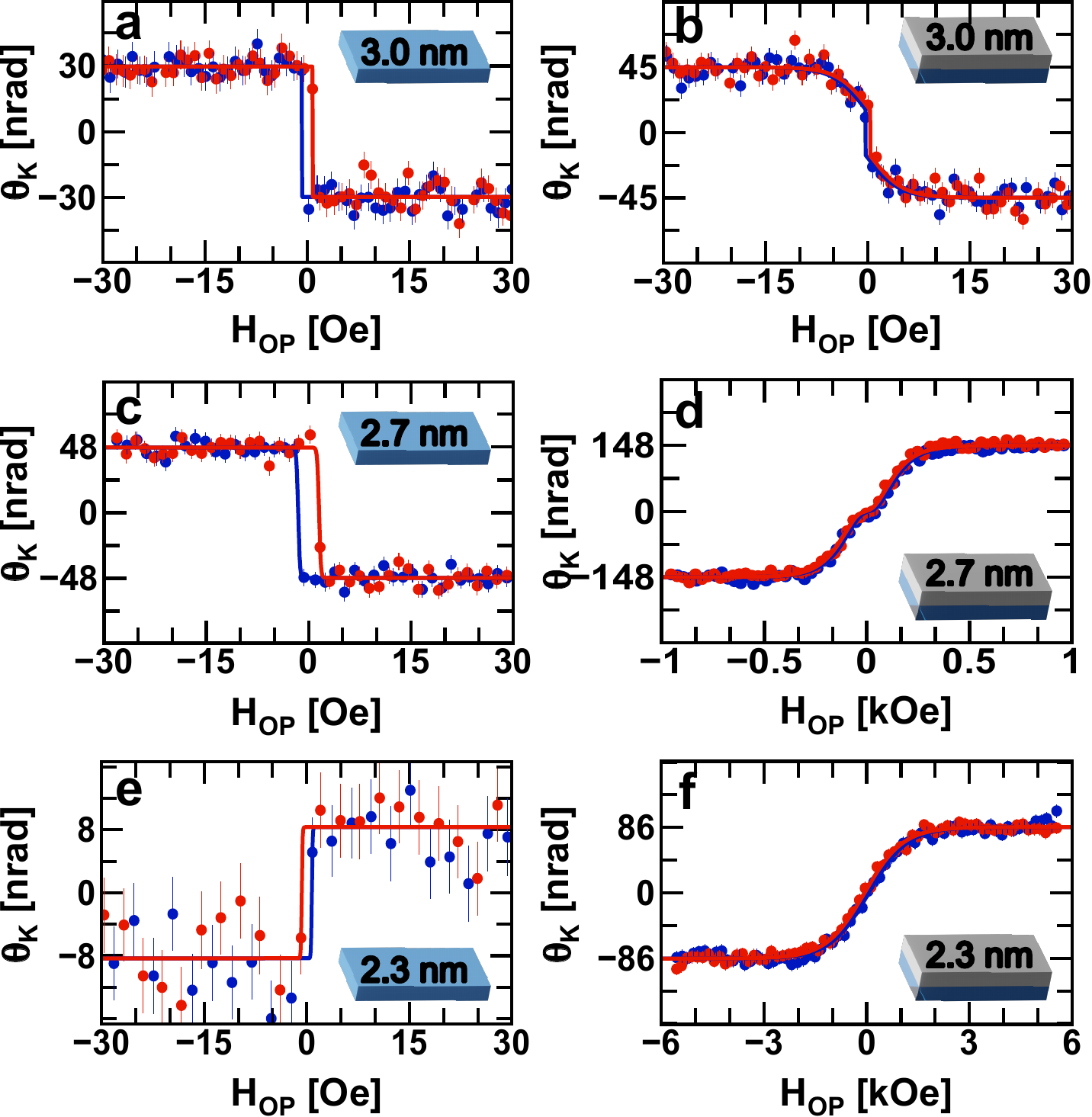}
        \caption{Polar SaMOKE data with an OP applied field taken from the bare TmIG (blue layer) of the thickness (a) 3 nm, (b) 2.7 nm, and (c) 2.3 nm, Data on the right column are taken from TmIG/Pt bilayers with a 2 nm thickness Pt layer (e.g. gray layer). Solid lines are fitted using methods described in the text. The OP experimental configuration is depicted in Fig. \ref{fig:app}d. Solid lines are fitted using methods described in the text. The left column shows measurements from the bare TmIG while the right column shows those for bilayers with a nominal 2 nm Pt over-layer}
        \label{fig:OP}
    \end{figure}

We first measured the samples with a sweeping OP field, $H_{OP}$, as depicted in Fig. \ref{fig:app}d. Figs. \ref{fig:OP}(a,b), (c,d), and (e,f) present measurements taken from the (TmIG,Pt/TmIG) regions of three samples with 3.0, 2.7, and 2.3 nm thick TmIG films, respectively. In other words, the left column includes measurements of bare TmIG films and the right column includes measurements of Pt/TmIG. The hysteresis in Figs. \ref{fig:OP}a, c, e, and f are well described by a single sigmoid, i.e. $a\tanh{(\frac{H-h_{0}}{\Delta H})}$ where $a$ is the Kerr angle at saturation, $H$ is the applied magnetic field, $h_{0}$ is the coercivity, and $\Delta H$ describes the magnetic hardness of the film. The hysteresis of the bilayers in Figs. \ref{fig:OP}b and d were not properly described by a single sigmoid, and a sum of two was needed, i.e. $a_{1}\tanh{(\frac{H-h_{0,1}}{\Delta H_{1}})}+a_{2}\tanh{(\frac{H-h_{0,2}}{\Delta H_{2}})}$, in order to achieve a reasonable fit. From fits of the data shown in Figs. \ref{fig:OP}a-f, we can extract coercive fields of $\approx$ 0.8, 0.3, 1.5, 8, 0.7, and 60 Oe respectively. These small coercive fields are consistent with the low damping of these TmIG films \cite{velez2019high,shao2018role,chanda2024temperature,lee2020interfacial}. The extraction of the saturation fields are discussed below. The parameters of all fits shown are given in the supplementary information.

We assume that all differences between the left and right columns arise from  the presence of the Pt/TmIG interface. 
All measurements in Figs. 2a, c, and e, have shown a hard PMA behavior and small switching fields with some notable differences. First, there is a conspicuous change in Kerr angle sign in the hysteresis data of the films. Figs. 2a and 2c (2e) give a negative (positive) Kerr angle for a positive applied field. Secondly, there is a lack of correlation of the Kerr angle magnitude at saturation with the film thickness. Since the Kerr angle is an optical observable arising from perturbations of a material’s optical constants from the spin-orbit interaction. The relative sign of the Kerr angle does not directly correspond to the magnetization direction, and there is no simple relation between the magnitude of the Kerr angle and the saturation magnetization. The direction of magnetization must be inferred from the direction of applied field. The sign change in Kerr angle most likely arises from an evolution of the TmIG electronic band structure. 

Although we observe a similar Kerr angle sign and magnitude differences in the two columns, the bilayer hysteresis shows a drastic dependence of the magnetic anisotropy as a function of TmIG thickness. This is evidenced by the large increase in the saturation field with decreasing thickness as well as an increase in the slope of the loop. The Pt/TmIG 3.0 nm bilayer data shows a soft perpendicular anisotropy. The softening of the anisotropy indicates that the interfacial anisotropy is nontrivial for this film thickness. The Pt/TmIG 2.7 nm bilayer shows a large increase in saturation field from the 3.0 nm bilayer. There is also a kink near zero field, a feature that we will elaborate more later. The Pt/TmIG 2.3 nm hysteresis has one order of magnitude larger saturation field with an elongated s-like curve that is the signature of a hard axis. These changes are evidence of an evolution from a perpendicular magnetic anisotropy to a compass anisotropy arising from an additional interfacial anisotropy at the Pt/TmIG interface \cite{banerjee2014enhanced,weber1997determining,berling2006accurate}. It is only in the ultrathin film limit that the effect of the Pt lay on the anisotropy becomes non-negligible. 

We estimate the saturation field to be the field for which the fitting function reaches 99\% of its asymptotic value. The uncertainty of the 99\% saturated field was determined numerically by assuming that all fitting parameters are uncorrelated. Each value of our fitting function variables has an associated standard error which corresponds to the standard deviation of a normal distribution of possible values for a given parameter. The calculated fit value and its corresponding standard error are used to generate a normal distribution of possible values with at least 50,000 instances. A Monte-Carlo method is then used to generate a normal distribution of 99\% saturation field values, the mean and standard deviation of which is listed in Table \ref{tab:saturation}. More details on the fitting functions and error analysis are given in the supplementary information.

\subsection{In-Plane Field Sweeps}

    \begin{figure}[t]
        \centering
        \includegraphics[width=\the\wid]{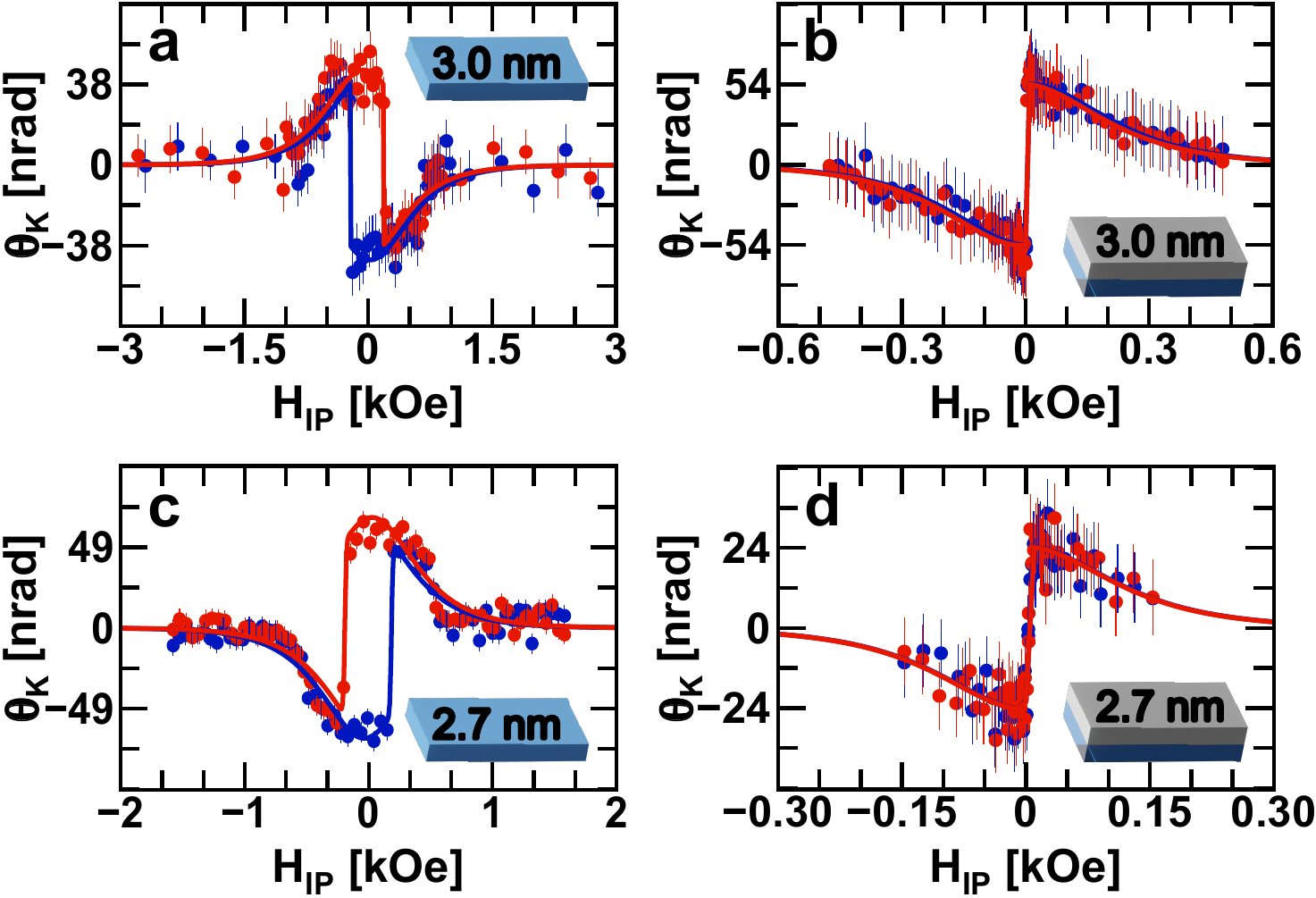}
        \caption{ Polar SaMOKE data with an IP applied field taken from (a) bare TmIG (3 nm), (b) Pt (2 nm)/TmIG (3 nm), (c) bare TmIG (2.7 nm), and (d) Pt (2 nm)/TmIG (2.7 nm). Solid lines are fitting curves using methods described in the text. The data was taken with the configuration depicted in Fig. \ref{fig:app}e. A linear paramagnetic background has been subtracted from the signal. }
        \label{fig:ip}
    \end{figure}

To quantify the interfacial anisotropy, we further perform MOKE measurements as a function of an in-plane field $H_{IP_{\parallel}}$ as depicted in Fig.~\ref{fig:app}e. The in-plane field is parallel to the bare TmIG and bilayer boundary. Figs. \ref{fig:ip} (a,b) and (c,d) represent the measurements from the (TmIG,Pt/TmIG) regions of the 3.0 and 2.7 nm TmIG films, respectively. The signal-to-noise ratio of the measurements on the 2.3 nm TmIG was not sufficiently large for the IP scans. The data in Fig. \ref{fig:ip}a and c show a slowly varying magnetization that is indicative of a hard axis magnetization rotation with a much larger saturation field than that found with an OP field. A similar trend in the bilayer data is shown in Figs. \ref{fig:ip}b and d. However, the perpendicular magnetic anisotropy is softened compared to the bare TmIG.  This provides further evidence that the presence of an interfacial anisotropy at the Pt/TmIG interface leads to an in-plane or compass anisotropy. A decreased saturation field is observed when the TmIG thickness changes from 3.0 to 2.7 nm in bilayers.

Fitting the IP data requires two assumptions. First, there are few domains in the area from which MOKE measurements are taken ($\sim$ 150 $\mu$m in diameter). In other words, the magnetization rotates with the applied magnetic field coherently. Second, there is a preferred plane of magnetization rotation that is defined by sample surface normal and the boundary between the Pt covered by bilayer and bare TmIG region. This plane (OP-IP$_{\parallel}$) depicted in Fig. \ref{fig:anisum} (green shaded plane) is defined by the shape anisotropy induced by the presence of the Pt partial film. The data in Fig. \ref{fig:ip} indicates that the first assumption is valid. Otherwise, there wouldn't be a well-defined OP magnetization component during the in-plane field sweep.

In the polar MOKE geometry (Fig. \ref{fig:app}d, e), only the OP magnetization was explicitly measured.  We derive a fitting function for the data in Fig. \ref{fig:ip}. 
We find that the $\sech()$ function gives a qualitatively correct result. However, it does not capture the discontinuities present in Fig. \ref{fig:ip}, where the Kerr angle suddenly flips sign. To account for these discontinuities, the $\mathrm{sech}()$ was multiplied by a $\tanh{((H-h_0)/\Delta H)}$ function. When $\Delta H$ approaches 0, $\tanh{((H-h_0)/\Delta H)}$ approximates a step function with the discontinuity at $h_0$. The saturation fields were determined from the fitting functions of all hysteresis measurements. The method for determining the saturation fields as well as the fitting functions of the hysteresis are described in the supplementary information. The 99\% saturation fields  and their estimated uncertainties are summarized in Table \ref{tab:saturation}. %using the same method described for the OP results above. 

%%%%%%%%%%%%%%%%%%%%%%%%%%%%%%%%%%%%%%%%%%%%%%%%%%%%%%%%%%%%%%%%%%
%%%%%%%%%%%%%%%%%%%%%%%%%%%%%%%%%%%%%%%%%%%%%%%%%%%%%%%%%%%%%%%%%%

\section{Discussion}

In addition to extracting the saturation fields from these films, we further analyze the change in anisotropy due to electronic effects at the Pt-TmIG interface. Magnetic proximity effect, i.e., induced magnetic moment in Pt due to TmIG, is assumed to be small and neglected. The difference between IP and OP saturation fields gives the effective anisotropy field that can be used to calculate the anisotropy energy density with the estimated TmIG magnetization. The magnetic anisotropy energy is a sum of both intrinsic and extrinsic effects. The intrinsic effects include exchange and crystal field interactions \cite{o2000modern}, and the extrinsic effects comprise of shape and strain-induced anisotropy. For example, the PMA in epitaxially grown TmIG and other garnets thin films typically originate from the extrinsic strain from the substrate (sGGG in our case)~\cite{rosenberg2021magnetic,vu2020tunable,lee2020engineering}. The change in anisotropy energy due to strain, $\Delta K_u$, for epitaxially grown TmIG (111) is separated into several terms  \cite{rosenberg2021magnetic,vu2020tunable}: 
\begin{equation}
    \Delta K_u = -\frac{K_1}{12} -\frac{1}{2}\mu_0 M_{s}^{2} + \frac{9}{4}\lambda_{111} c_{44} (\frac{\pi}{2}-\beta)
    \label{eq:epianis}
\end{equation}
where $K_1$ is the cubic magnetic anisotropy constant, $M_s$ is the saturation magnetization, $\lambda_{111}$ is the magnetostriction coefficient, $c_{44}$ is the shear stiffness constant, and $\beta$ is the angle describing the shear distortion away from cubic symmetry. The epitaxially induced change in anisotorpy happens thorughout the whole volume of the film. Therefore, the anisotropy energy density is independent of film thickness and should be the same for all samples measured. Consequently,  the bare TmIG films maintain their PMA for all thicknesses as observed in Figs. \ref{fig:OP}a, c, and e. %The change of the magnetic anisotropy due to the epitaxial strain does not depend on the film thickness, it is constant in film volume. 

\begin{table}[t]
    \centering
    \begin{tabular}{|c||c|c|c|}
        \hline
        \begin{tabular}{@{}c@{}}Field \\ Orientation\end{tabular} & $t_{\scaleto{TmIG}{4pt}}$ & Coverage  & $ H_{sat}$ \\% & $\theta_{K}^{sat}$\\
        \hline
        \hline
        \multirow{6}{*}{ OP}   & \multirow{2}{*}{ 3.0 nm} & Pt & ~~11.3 $\pm$ 3.1 Oe~~ \\\cline{3-4}
                                &                       & None & ~~0.83 $\pm$ 0.11 Oe~~\\\cline{2-4}
                                & \multirow{2}{*}{ 2.7 nm} & Pt &  ~~308 $\pm$ 69 Oe~~\\\cline{3-4}
                                &                       & None &  ~~2 $\pm$ 17 Oe~~\\\cline{2-4}
                                & \multirow{2}{*}{ 2.3 nm} & Pt &  ~~3.04 $\pm$ 0.15 kOe~~\\\cline{3-4}
                                &                       & None &  ~~8.3 $\pm$ 0.8 Oe~~\\
        \hline
             
        \multirow{4}{*}{ IP$_{\parallel}$}       & \multirow{2}{*}{ 3.0 nm}& Pt &  ~~0.89 $\pm$ 0.03 kOe~~ \\\cline{3-4}
                                &                      & None &  ~~2.22 $\pm$ 0.13 kOe~~ \\ \cline{2-4}
                                & \multirow{2}{*}{ 2.7 nm}& Pt & ~~0.49 $\pm$ 0.23 kOe~~ \\\cline{3-4}
                                &                      & None & ~~1.61 $\pm$ 0.09 kOe~~ \\
        \hline
    \end{tabular}
    \caption{Saturation fields extracted from hysteresis fits. The fields are estimated from fitting functions using the 99\% of asymptotic value criteria}%\notetn{Bare->TmIG and Pt->bilayer, change column width}}
    \label{tab:saturation}
\end{table}

In contrast, the corresponding bilayer data shown in the right column of Fig. \ref{fig:OP} indicates  an evolution of the anisotropy when the TmIG film thickness is 3.0 nm and below. The perpendicular magnetic anisotropy is maintained in a 3.5 nm Pt/TmIG bilayer \cite{nunley2022quantifying} (Data included in Supplementary Information). 
Thus, we infer that the transition from strong to soft PMA occurs between 3.5 and 3.0 nm for Pt-covered TmIG. In this range, the interfacial anisotropy originating from Pt and TmIG interface becomes significant compared to other sources (e.g. shape anisotropy or the strain between the substrate and TmIG). While other sources of magnetic anisotropy scale with the volume of the film, the interfacial anisotropy scales with the interfacial area. 

We can add the volumetric and interfacial terms after dividing the interfacial anisotropy energy density by the film thickness. The sum is written as 
\begin{equation}
    K_{tot}(\theta) = K_{vol}\sin^2(\theta)^2 + K_{int}/t_{\scaleto{TmIG}{4pt}}\sin^2(\theta+\frac{\pi}{2})^2
    \label{eq:anisvolint}
\end{equation}
In this expression, $K_{tot}$, $K_{vol}$ and $K_{int}$ refers to the total, volumetric, and interfacial anisotropy, respectively. $t_{\scaleto{TmIG}{4pt}}$ is the film thickness, and $\theta$ describes the angle between the magnetization and surface normal (OP). Both $K_{vol}$ and $K_{int}$ have cylindrical symmetry reflected in their dependence in $\theta$.  Eq. \ref{eq:epianis} includes several important sources of  $K_{vol}$ but other effects (e.g. shape anisotropy) may also contribute. 

While $K_{vol}$ is responsible for the net PMA in the bare films, the competition between the volumetric and interfacial terms causes the transition from a perpendicular magnetic anisotropy to a compass anisotropy. This is reflected in Eq. \ref{eq:anisvolint} with the second interfacial term, $K_{int}/t_{\scaleto{TmIG}{4pt}}$ asymptotically approaching zero with increasing film thickness, $t_{\scaleto{TmIG}{4pt}}$. For the film thicknesses in this work, $t_{\scaleto{TmIG}{4pt}}$ is small and therefore the $K_{int}/t_{\scaleto{TmIG}{4pt}}$ term is large. We can isolate the effect of the $K_{int}/t_{\scaleto{TmIG}{4pt}}$ in Eq. \ref{eq:anisvolint} as effective anisotropy fields calculated using the saturation fields given in Table \ref{tab:saturation}.

Table \ref{tab:intanis} summarizes the effective IP$_{\parallel}$/OP anisotropy fields for the 3.0 and 2.7 nm samples. In this table, the $H_{anis}^{eff}$ is defined as the difference between the saturation fields from Table \ref{tab:saturation}, i.e., $H_{anis}^{eff}=H_{sat}^{IP_{\parallel}} - H_{sat}^{OP}$. $H_{anis}^{int}$ is the difference in the effective anisotropy fields for the TmIG films with and without Pt, corresponding to the interfacial term in Eq. \ref{eq:anisvolint}. %$H_{anis}^$ the effective IP$_{\parallel}$/OP anisotropy fields for the the 3.0 and 2.7 nm samples. 
The largest anisotropy field is found in the 3.0 nm TmIG film. The addition of Pt reduces the anisotropy field by 1340 Oe, resulting in an effective anisotropy field that is less than half that of the bare TmIG film. This decrease indicates that the interfacial anisotropy field is (1340 Oe)$\cdot$(3.0 nm)~$\approx$~ 4.0 kOe$\cdot$nm for the 3.0 nm film and similarly 1.4 kOe$\cdot$(2.7 nm)~$\approx$~3.8 kOe$\cdot$nm for the 2.7 nm film.

We can now  quantify the anisotropy energy density at the Pt-TmIG interface using $H_{anis}^{int}$ in Table \ref{tab:intanis}. By assuming a magnetization of ~115 emu/cm$^{3}$, corresponding to the TmIG sputtering target, we convert the interfacial anisotropy field to an energy density, $K_{int}/t_{\scaleto{TmIG}{4pt}} = H^{int}_{anis} \cdot M_{sat}$. We find interfacial anisotropy energy densities, $K_{int}$, of 46 $\pm$ 6 and 44 $\pm$ 13 $[\mu J/m^{2}]$ for the 3.0 and 2.7 nm bilayers respectively as listed in Table \ref{tab:intanis}. 

~Given the uncertainties of the extracted interfacial anisotropies, these values are in reasonable agreement as expected: the interfacial area energy density should not be dependent on film thickness. The origin of this interfacial anisotropy is most likely Rashba spin-orbit coupling \cite{lee2020interfacial,banerjee2014enhanced} which are also expected to produce an interfacial DMI.

\begin{table}[t]
\begin{tabular}{|c|c|c|c|c|}
    \hline
    \multicolumn{1}{|c|}{$t_{\scaleto{TmIG}{4pt}}$} & \multicolumn{1}{c|}{Coverage} & \multicolumn{1}{c|}{$H_{anis}^{eff}$ [Oe]} & \multicolumn{1}{c|}{$H_{anis}^{int}$ [Oe]} & \multicolumn{1}{c|}{$K_{int} [\mu J/m^{2}]$}  \\ \hline
    \hline
    \multirow{2}{*}{ 3.0 nm}  & None & ~2219 $\pm$ 130~ & \multirow{2}{*}{~1340 $\pm$ 160~} & \multirow{2}{*}{46 $\pm$ 6} \\ \cline{2-3} 
                              &  Pt  & ~879 $\pm$ 33~  &   & \\ 
    \hline
    \multirow{2}{*}{ 2.7 nm}  & None & ~1608 $\pm$ 107~ & \multirow{2}{*}{~1430 $\pm$ 410~} & \multirow{2}{*}{44 $\pm$ 13} \\ \cline{2-3} 
                              & Pt   & ~182 $\pm$ 299~   &  & \\ \hline
\end{tabular}
\caption{Effective anisotropy fields calculated from the difference in saturation fields. There is a large decrease in anisotropy field with the addition of the Pt film. The interfacial anisotropy energy density found for both TmIG thicknesses are consistent after removing the thickness dependence assuming that the magnetization per unit volume, $M_{s}$, is the same for each sample. }
\label{tab:intanis}
\end{table}

    \begin{figure}[t!]
        \centering
        \includegraphics[width=\the\wid]{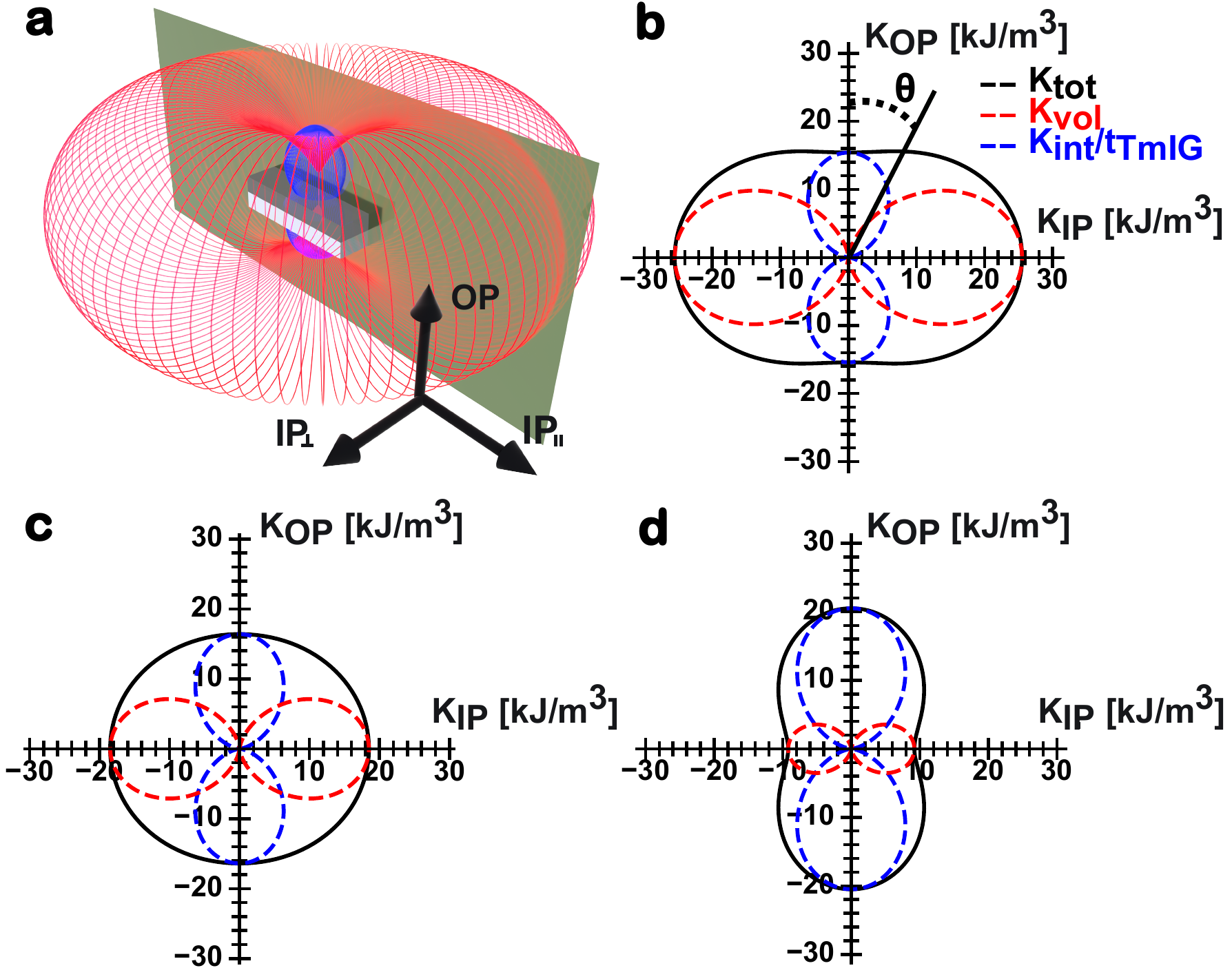}
        \caption{ Anisotropy energy density surfaces and cross-sections. (a) A 3D rendering of the magnetic anisotropy energy density surface of the sample 3.0 nm bilayer sample with OP uniaxial anisotropy energy density surface (red) and IP$_{\parallel}$ uniaxial anisotropy energy density surface (blue) shown. The green plane indicates the plane of magnetization rotation for the data shown in Fig. \ref{fig:ip}. Energy density surface cross-sections with the green plane for (b) 3.0 nm, (c) 2.7 nm, and (d) 2.3 nm Pt/TmIG bilayers, respectively.   $K_{tot}$ (black), $K_{vol}$ (red), and $K_{int}/t$ (blue)  are representative of terms in Eq. \ref{eq:anisvolint}. (b) shows the direction from which $\theta$ is measured as indicated in Eq. \ref{eq:anisvolint}.} 
        \label{fig:anisum}
    \end{figure}

The evolution of the volumetric and interfacial anisotropy competition as a function of TmIG thickness can be visualized in Fig. \ref{fig:anisum}. The anisotropy energy density cross-sections (with the plane indicated by green) from 3.0, 2.7, and 2.3 nm Pt/TmIG bilayers are plotted in Figs.~\ref{fig:anisum}b, c, and d, respectively. The red, blue, and black curves correspond to the volumetric, interfacial anisotropy, and their sum, i.e. the net anisotropy energy density, respectively. Figs. \ref{fig:anisum}b and c depict the results summarized in Tables \ref{tab:saturation} and \ref{tab:intanis}. Fig. \ref{fig:anisum}d shows the expected energy surface of the 2.3 nm bilayer assuming an anisotropy energy density given by the average $H_{anis}^{eff}$ of the 3.0 and 2.7 nm bare films and the same $H_{anis}^{int}*t_{\scaleto{TmIG}{4pt}}$. Fig. \ref{fig:anisum}b shows the dominant uniaxial anisotropy is OP for the 3 nm Pt/TmIG bilayer. In contrast, Fig. \ref{fig:anisum}d indicates the opposite scenario with an IP easy plane in 2.3 nm thick Pt/TmIG bilayer. In  2.7 nm thick Pt/TmIG bilayer, the IP$_{\parallel}$ and OP uniaxial anisotropies are approximately equal. There is no dominant easy or hard axis. In other words, a nearly isotropic magnet as depicted in Fig. \ref{fig:anisum}c is found. Although this picture describes the evolution well of magnetic anisotropy well, it doesn't explain a plateau near the zero applied field observed in the hysteresis of the 2.7 nm Pt/TmIG bilayer shown in Fig. \ref{fig:OP}d. Micromagnetic simulations (details in the supplementary information) indicate that this feature may arise from a combination of interfacial DMI and weak net anisotropy.

\section{Conclusions}

In summary, we used a Sagnac interferometer to extend the characterization of magnetic properties to ultrathin TmIG films. By performing MOKE experiments on ultrathin bare TmIG and Pt/TmIG  bilayers and comparing scans with an applied magnetic field with OP and IP directions, we quantify the Pt/TmIG anisotropy energy density via the changes in the extracted saturation fields. While the PMA is established by epitaxial strain between the TmIG/sGGG substrate, the compass anisotropy in the bilayers originates from the Pt/TmIG interface. We confirm that epitaxial strain-induced PMA is maintained down to 2.3 nm thick TmIG films. We find that the effective anisotropy field reduces by more than 25\% in a bare TmIG film when its thickness reduces from 3.0 to 2.7 nm. Even in 3.0 nm thick TmIG, the presence of a 2 nm thick Pt layer reduces the effective anisotropy field by more than a factor of 2. Our studies provide valuable information to design spintronic devices based on ultrathin ferrimagnetic insulating films and bilayers. \\\\

\textit{Acknowledgements} T. N. Nunley thanks James Erskine for helpful conversations. This research was initiated and primarily supported by a collaborative grant (Li and Yang) from NSF EPM program grant DMR-2225645. X. Li gratefully acknowledges funding from the Center for Energy Efficient Magnonics an Energy Frontier Research Center funded by the U.S. Department of Energy, Office of Science, Basic Energy Sciences at SLAC National Laboratory under contract DE-AC02-76SF00515 (low-damping magnetic materials) and the Welch Foundation Chair F-0014 (materials supply). L. J. Chang while visiting UT-Austin was provided by a Portugal-UT collaboration grant. The collaboration between S,F. Lee and X. Li is facilitated by the AFOSR grant FA2386-21-1-4067. Guo acknowledges support (film growth) by the Center for Emergent Materials, an NSF MRSEC, under Grant No. DMR-2011876.

% \nocite{*}
\bibliography{strain}% Produces the bibliography via BibTeX.

\end{document}